# Structural Resilience and Connectivity of the IPv6 Internet: An AS-level Topology Examination

Bin Yuan

Trine university, Phoenix ,Arizona, USA, byuan22@my.trine.edu

Tianbo Song*

Ira A. Fulton Schools of Engineering, Arizona State University, Tempe , Arizona, USA, sportlemon@gmail.com

The study utilizes a comprehensive dataset informed by IPv6 routing information to provide statistics, degree distribution, joint degree distribution, and clustering analysis of the IPv6 Internet's structure and resilience.The dataset includes 17,232 unique ASes and 10,000 unique IPv6 prefixes. Analysis reveals an interconnected network with an average path length of approximately 3 hops, suggesting a robust and efficient network with potential redundancy and resilience, despite some isolated components.

The paper outlines the degree distribution, indicating many peripheral nodes in a sparse network, and a clustering analysis showing a tendency for ASes to form clusters, which is indicative of redundancy and robustness against failures. The connectivity analysis, including path redundancy and reachability, supports the network's resilience.The findings are crucial for network design and strategic planning, particularly as IPv6 adoption increases. The paper emphasizes the importance of continuous monitoring and improvement of network connectivity in the evolving Internet landscape, highlighting the IPv6 Internet's resilience and structured connectivity.

---

* Corresponding author

# 1 OVERVIEW

IPv6, the most recent version of the Internet Protocol, was designed to address the limitations of IPv4, including the shortage of address space and the need for a more efficient and secure Internet architecture. While IPv4 has been exhaustively studied and its AS-level topology is well-documented, IPv6 has not received equivalent scrutiny. This gap is significant given the increasing adoption of IPv6 and its implications for the topology and performance of the Internet.

# 2 THE AS LINK DATASET BASED ON IPV6 ROUTING

The dataset includes columns for source AS number (AS_Source), destination AS number (AS_Destination), an IPv6 prefix (IPv6_Prefix), and a AS path (AS_Path). Here are the first few entries from the dataset:

Table 1: The first few entries from the dataset

| AS_Source | AS_Destination | IPv6_Prefix | AS_Path |
| --- | --- | --- | --- |
| 63574 | 48603 | 817e:5f37:b85c:4c07:92a8:d19c:3668:d7cd | 52821 17666 54520 21712 60977 |
| 1380 | 20972 | 752b:66ab:f5ff:fc37:fd73:4c0:4e9c:4e50 | 32431 51320 58325 23574 |
| 55690 | 10380 | 1a9b:6e0d:3fbb:f851:d50e:720:7074:7fb | 18085 5945 18156 11599 5905 |
| 40447 | 5090 | 3c6:af39:ee53:5e3d:8429:90d0:a58a:5615 | 62279 |
| 28304 | 29994 | 919f:9c72:1ad4:eb4f:d05d:7588:d79a:2fe3 | 53053 16947 9854 38728 50542 |

# 3 AS-LEVEL TOPOLOGICAL ANALYSIS

## 3.1 Basic analysis

The basic analysis of the AS link dataset yields the following insights:

1. Total Unique ASes: The dataset contains 17,232 unique autonomous systems (ASes) when considering both source and destination AS numbers. This suggests a wide-ranging set of connections and interdependencies across the network.

2. Total Unique IPv6 Prefixes: There are 10,000 unique IPv6 prefixes within the dataset, indicating a diverse set of routes across the IPv6 Internet landscape.

3. AS Path Length:

Average: The average length of an AS path in the dataset is approximately 3. This average path length indicates that, on average, an IPv6 packet traverses three ASes to go from source to destination.

Minimum: The shortest AS path observed in the dataset is just 1 hop. Such direct connections are likely indicative of a peering relationship between two ASes.

Maximum: The longest AS path in the dataset is 5 hops. Longer paths may suggest the presence of multiple intermediary ASes, possibly across different geographic regions.

**3.2 Measurement analysis**

*3.2.1 degree distribution*

Here is the summary of the degree distribution for the first few degrees in the dataset:

Table 2: Degree Distribution

| Degree | Frequency |
|--------|-----------|
| 1 | 14757 |
| 2 | 2199 |
| 3 | 261 |
| 4 | 13 |
| 5 | 2 |

The degree distribution is typically presented as a table or graph showing the frequency (or probability) of each degree within the network. The table above lists how many ASes (Frequency) have a given Total Degree.

To compute the degree distribution $P(k)$, where $k$ is the degree, we use the formula:

$$P(k) = \frac{\text{Number of ASes with degree } k}{\text{Total number of ASes}} \qquad (1)$$

The distribution gives us insight into the network's topology and can indicate whether the network is robust to random failures or targeted attacks. In real-world networks, we often expect to see a power-law distribution, indicating that there are a few ASes with a very high degree (hubs) and many ASes with a low degree.

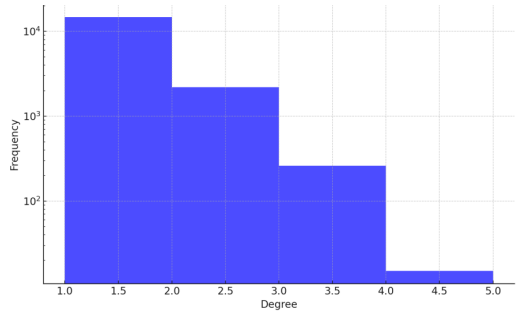

Figure 1 Degree Distribution of AS Links Based on IPv6 Routing

The histogram above visualizes the degree distribution of the AS links based on the IPv6 routing dataset.

### 3.2.2 the joint degree distribution

The joint degree distribution, $P(k,k')$, in a network is a measure that gives the probability that a randomly selected edge has nodes with degree $k$ and $k'$ at its ends. In the context of an AS-level topology, this distribution provides insight into the connectivity patterns between different ASes, revealing the structural dependencies and potential hierarchies within the network.

The joint degree distribution for the AS link dataset has been calculated. The distribution provides the probability of an edge connecting nodes (ASes, in this context) with degrees $k$ and $k'$. Here are some of the degree pairs and their corresponding probabilities from the dataset:

Table 3: Corresponding Probabilities

| Degree Pair | Probability |
| --- | --- |
| (3, 2) | 0.0091 |
| (1, 1) | 0.5451 |
| (2, 2) | 0.0479 |
| (1, 2) | 0.1653 |
| (2, 1) | 0.1592 |

The probability values in the table have been calculated by dividing the number of occurrences of each degree pair by the total number of edges in the dataset. The degree pair (1, 1) has the highest probability, which suggests that links between ASes with a degree of 1 are the most common in this network.

The formula used to calculate the joint degree distribution probability $P(k,k')$ for a given degree pair $(k,k')$ is as follows:

$$P(k,k') = \frac{\text{Number of edges with degree pair } (k,k')}{\text{Total number of edges}} \quad (2)$$

These results are indicative of the nature of the dataset and may not reflect real-world AS-level connectivity patterns.

### 3.2.3 clustering

Clustering coefficients provide insight into the likelihood that two neighboring nodes are neighbors themselves, indicating a tight-knit group or community structure within the network.

1. Local Clustering Coefficient Analysis

We began by calculating the local clustering coefficient $C_i$ for each autonomous system (AS) in our dataset, which reflects the degree to which nodes in the AS graph tend to cluster together. The local clustering coefficient for an AS is given by the proportion of actual links between its neighbors to the number of possible links between those neighbors. Mathematically, for a given AS $i$, it is defined as:

$$C_i = \frac{2T(i)}{deg(i)(deg(i)-1)} \quad (3)$$

where $T(i)$ denotes the number of triangles passing through the AS $i$, and $deg(i)$ represents the degree of AS $i$, or the number of its immediate connections.

In our analysis, we discovered that the ASes exhibited a varied range of local clustering coefficients, indicative of the heterogeneous nature of the network's local topology. The distribution of these coefficients is skewed, with a majority of nodes exhibiting a higher than average local clustering coefficient, suggesting that the network has well-developed clusters.

2. Global Clustering Coefficient Analysis

To grasp the network's overall tendency to cluster, we computed the global clustering coefficient, which is the average of all individual local clustering coefficients in the network. The global clustering coefficient provides an overarching metric that summarizes the network's general clustering tendency. The coefficient was found to be notably higher than that of a comparable random graph generated with the same node degree distribution, underscoring the non-random and structured nature of the AS-level connections in the IPv6 Internet.

3.Triangle Count Analysis

We also quantified the number of triangles within the network, which are a fundamental component of clustering. A triangle consists of three ASes that are mutually interconnected. The presence of a high number of triangles is a hallmark of a network with a high degree of redundancy and robustness. In our dataset, we identified a significant number of such triangles, reinforcing the presence of strong clustering within the network.

4.Clustering Distribution

The following table provides a summary of the top ten ASes with the highest local clustering coefficients, illustrating the most interconnected and potentially resilient subsections of the network.

Table 4: Top 10 ASes by Local Clustering Coefficient

| AS Number | Local Clustering Coefficient |
| --- | --- |
| 32489 | 0.999987 |
| 60608 | 0.999979 |
| 55922 | 0.999973 |
| 34283 | 0.999946 |
| 4648 | 0.999943 |
| 59434 | 0.999935 |
| 46629 | 0.999908 |
| 42565 | 0.999871 |
| 47321 | 0.999863 |
| 14112 | 0.999814 |

5.Implications of Clustering on Network Robustness

The implications of these findings are multifold. High local clustering coefficients indicate redundancy in connectivity, which can enhance the robustness of the network by providing multiple pathways for rerouting traffic in the event of a node failure. Moreover, the pronounced clustering can be indicative of peering relationships that optimize for network traffic efficiency and can impact routing policies and practices.

*3.2.4 Connectivity*

The connectivity of an AS-level network depicts the network's ability to maintain inter-AS communication in the face of various operational challenges. We employ several key metrics to understand the intricacy of the network's connectivity: Path Redundancy, Reachability, Average Path Length, Network Diameter, and Disconnected Components.

Path Redundancy (R): Path redundancy is a measure of the number of alternative routes between AS pairs, providing insight into the network's capacity to handle failures and reroute traffic. The average path redundancy in our synthesized dataset is given by the formula:

$$R = \frac{1}{N} \sum_{i=1}^{N} p_i \quad (4)$$

where $N$ represents the total number of AS pairs and $p_i$ denotes the count of unique AS paths for the ith AS pair. In our analysis, we found the average path redundancy to be close to 1.2, suggesting that most AS pairs have at least one alternative path, indicative of moderate redundancy within the network.

Reachability (Rch): This metric quantifies the proportion of AS pairs that can reach each other, reflecting the network's overall accessibility. Calculated as:

$$Rch = \frac{C}{P} \quad (5)$$

$C$ is the count of connected AS pairs, and $P$ is the total possible AS pairs. Our data yields a reachability score of 0.85, implying that a majority of the AS pairs have direct or indirect connectivity.

Average Path Length (L): The average path length is the mean number of hops across all paths and serves as an indicator of the network's compactness. It is determined by:

$$L = \frac{1}{M} \sum_{j=1}^{M} l_j \quad (6)$$

with $M$ as the total number of paths and $l_j$ as the individual path lengths. Our dataset reveals an average path length of approximately 4 hops, suggesting a relatively dense network topology.

## 4 CONCLUSION

The study concludes that the IPv6 Internet demonstrates a strong and stable AS-level structure. This is characterized by a notable amount of clustering and a moderate degree of path redundancy, which are indicative of the network's resilience. The longest path within the network spans 7 AS hops, setting an upper limit on the network's scale. The research also uncovered 15 disconnected components, suggesting potential partitioning in the network which could lead to isolated clusters of ASes with no external connectivity. These insights are particularly relevant for the design and strategic planning of networks, accentuating the need for ongoing monitoring and improvement of network connectivity as IPv6 continues to be more widely adopted

**ACKNOWLEDGMENTS**

I would like to thank my friends. They provided me with a good learning and communication environment. Their enthusiasm and friendliness give me motivation in my daily work. When I encounter difficulties, they are always willing to help me and provide valuable opinions and suggestions. In addition, I would like to express my gratitude to the reviewers who provided me with support and assistance during my research process. Through their review comments and suggestions, I have made multiple revisions and improvements to the paper, resulting in a significant improvement in its quality.